\documentclass[
 aip,
 amsmath,
 amssymb,
 reprint,
]{revtex4-1}

\usepackage{enumerate}
\usepackage{newtxtext}
\usepackage{booktabs}
\usepackage{dcolumn}
\usepackage{graphicx}
\usepackage{multirow}
\usepackage[colorlinks=true, citecolor=blue, urlcolor=blue, linkcolor=blue]{hyperref}
\usepackage[version=4]{mhchem}
\usepackage[dvipsnames]{xcolor}

\newcommand{\Vcal}[0]{\mathcal{V}}
\newcommand{\Ncal}[0]{\mathcal{N}}
\newcommand{\abs}[1]{\left\lvert#1\right\rvert}

\newcommand{\avg}[1]{\left\langle#1\right\rangle}
\DeclareMathOperator*{\argmin}{\arg\!\min}

\newcommand{\centered}[1]{\begin{tabular}{l} #1 \end{tabular}}

\let\originalleft\left
\let\originalright\right
\renewcommand{\left}{\mathopen{}\mathclose\bgroup\originalleft}
\renewcommand{\right}{\aftergroup\egroup\originalright}

\makeatletter
\newenvironment{tsubarray}[1]{%
  \vcenter\bgroup
  \Let@ \restore@math@cr \default@tag
  \baselineskip\fontdimen10 \scriptfont\tw@
  \advance\baselineskip\fontdimen12 \scriptfont\tw@
  \lineskip\thr@@\fontdimen8 \scriptfont\thr@@
  \lineskiplimit\lineskip
  \check@mathfonts
  \ialign\bgroup\ifx c#1\hfil\fi
    \normalfont\fontsize\sf@size\z@\selectfont\ignorespaces##\unskip\hfil\crcr
}{\crcr\egroup\egroup}
\makeatother
\newcommand{\tsub}[1]{\begin{tsubarray}{l}#1\end{tsubarray}}

\begin{document}


\title{A local orientational order parameter for systems of interacting particles}

\author{John \c{C}amk{\i}ran}
\email[]{john.camkiran@utoronto.ca}
\affiliation{\mbox{Department of Materials Science and Engineering, University of Toronto, Toronto, Ontario, M5S 3E4, Canada}}

\author{Fabian Parsch}
\affiliation{\mbox{Department of Mathematics, University of Toronto,  Toronto, Ontario, M5S 2E4, Canada}
}
\affiliation{\mbox{Department of Materials Science and Engineering, University of Toronto, Toronto, Ontario,  M5S 3E4, Canada}}

\author{Glenn D. Hibbard}
\affiliation{\mbox{Department of Materials Science and Engineering, University of Toronto, Toronto, Ontario, M5S 3E4, Canada}}

\date{\today}

\begin{abstract}
Many physical systems are well modeled as collections of interacting particles. Nevertheless, a general approach to quantifying the absolute degree of order immediately surrounding a particle has yet to be described. Motivated thus, we introduce a quantity $E$ that captures the amount of pairwise informational redundancy among the bonds formed by a particle. Particles with larger $E$ have less diversity in bond angles and thus simpler neighborhoods. We show that $E$ possesses a number of intuitive mathematical properties, such as increasing monotonicity in the coordination number of Platonic polyhedral geometries. We demonstrate analytically that $E$ is, in principle, able to distinguish a wide range of structures and conjecture that it is maximized by the icosahedral geometry under the constraint of equal sphere packing. An algorithm for computing $E$ is described and is applied to the structural characterization of crystals and glasses. The findings of this study are generally consistent with existing knowledge on the structure of such systems. We compare $E$ to the Steinhardt order parameter $Q_6$ and polyhedral template matching (PTM). We observe that $E$ has resolution comparable to $Q_6$ and robustness similar to PTM despite being much simpler than the former and far more informative than the latter.
\end{abstract}

\maketitle


\section{Introduction}
\label{sec:introduction}

Much can be said about the microscopic dynamics and macroscopic properties of a system given its structure---the spatial relationships between its constituents. It is interesting therefore that the general problem of indicating the structure of the immediate surrounding of a particle still lacks a satisfactory solution. Such a solution would comprise a single nonnegative real number that increases with the absolute degree of order immediately surrounding that particle---\textit{a local orientational order parameter}. While many indicators of local structure have been discussed,\cite{stukowski_2012, tanaka_2019} the vast majority of them are not order parameters in this sense. Instead, most existing indicators constitute \textit{classifiers} of local structure, which have great practical value in making nominal distinctions between the geometries of particle neighborhoods. Such indicators, however, have limited theoretical value, as they provide little, if any, insight into the geometric foundations of ordering in physical systems. Conversely, the few local structural indicators that are order parameters in the above sense are either ineffective or inefficient in making basic crystallographic distinctions.

\begin{figure}[!b]
    \includegraphics[width=\linewidth]{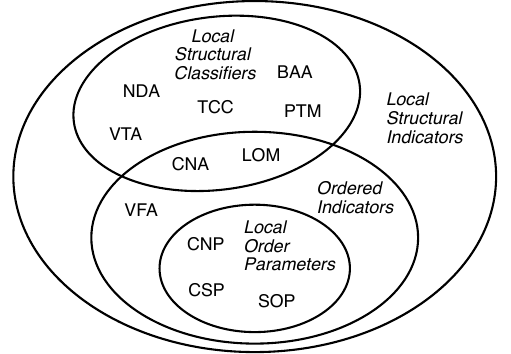}
    \caption{Euler diagram of common approaches to studying local structure: Voronoi face analysis (VFA),\cite{bernal_1959, tanemura_1977} Steinhardt order parameters (SOP),\cite{steinhardt_1983} common neighbor analysis (CNA),\cite{honeycutt_1987} centrosymmetry parameter (CSP),\cite{kelchner_1998} bond angle analysis (BAA),\cite{ackland_2006} common neighborhood parameter (CNP),\cite{tsuzuki_2007} neighbor distance analysis (NDA),\cite{stukowski_2012} topological cluster classification (TCC),\cite{malins_2013_b} Voronoi topology analysis (VTA),\cite{lazar_2015} polyhedral template matching (PTM),\cite{larsen_2016} and local order metric (LOM).\cite{martelli_2018}
    }
    \label{fig:euler_diagram}
\end{figure}

In this work, we introduce a local orientational order parameter that is both effective and efficient in distinguishing between the geometries of particle neighborhoods. The conception of such an indicator involves two main challenges: (I) the robust demarcation of particle neighborhoods and (II) the absolute quantification of local orientational order. We address the former challenge with a stochastic adjustment to the Voronoi tessellation. Briefly, the adjustment omits from a Voronoi neighborhood those particles whose inclusion in the neighborhood is sensitive to small changes in particle positions. We address the latter challenge with a simple coefficient that captures the amount of informational redundancy in the geometry of the immediate surrounding of a particle. It achieves this by comparing the number of pairs of bonds involving a particle to the number of different angles made by these pairs. Such a quantity appears to be able to both rank a particle neighborhood based on its absolute degree of orientational order and classify its geometry. To the best of our knowledge, it is the first structural indicator that brings together these qualities (see \mbox{Fig. \ref{fig:euler_diagram}} for a graphical survey of literature).

The remainder of this work is organized as follows: \mbox{Sec. \ref{sec:demarcating_neighborhoods}} discusses the demarcation of particle neighborhoods. \mbox{Sec. \ref{sec:quantifying_order}} concerns the quantification of local orientational order. \mbox{Sec. \ref{sec:computation}} describes an algorithm for computing our indicator. \mbox{Sec. \ref{sec:validation}} covers its numerical validation. \mbox{Sec. \ref{sec:application}} considers applications. Finally, \mbox{Sec. \ref{sec:discussion}} reflects on our results and expresses concluding remarks.

\section{Demarcating particle neighborhoods}
\label{sec:demarcating_neighborhoods}

\begin{figure}[!b]
    \centering
    \includegraphics{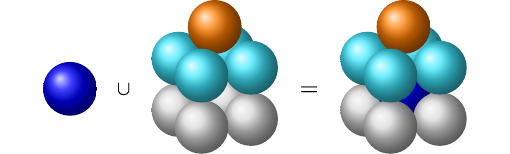}
    \caption{A particle and its neighborhood compose a central cluster.}
    \label{fig:central_cluster}
\end{figure}

We begin with a few key definitions. Let $S$ be a discrete set of points in $D$-dimensional Euclidean space. Call $S$ a \textit{system}, its every element $p$ a \textit{particle}, the particles $q_i$ adjacent to $p$ its \textit{neighbors}, and any set of adjacent particles in $S$ a \textit{cluster}. Together, a particle $p$ and its neighbors $q_i$ compose a cluster of the kind illustrated in Fig. {\ref{fig:central_cluster}}. Of course not every cluster possesses this particle-neighborhood form. We term those that do \textit{central} clusters and those that do not \textit{noncentral} clusters. Some works on local structural indication are agnostic to this distinction.{\cite{malins_2013_a, malins_2013_b}} Here, we argue that a truly local characterization of structure must unambiguously correspond to a locality of the system, that is, a particle. This is visibly not the case with noncentral clusters, and so, they are not considered in this work.

\begin{figure}[!t]
    \centering
    \includegraphics{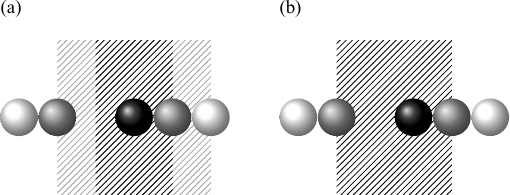}
    \caption{(a) Naive neighborhood of the central particle for ${\tau = 0}$ (dark hatched region) and ${\tau = 1}$ (all hatched regions). (b) Voronoi neighborhood of the same. Under the naive model, it is visible that no choice of $\tau$ can include adjacent particles (in gray) while excluding nonadjacent ones (in white). The Voronoi model, meanwhile, readily achieves this result.}
    \label{fig:neighborhood_models}
\end{figure}

The neighborhood of a particle is a qualitative concept with no unequivocal quantitative counterpart but rather a few discrete models. Perhaps the simplest of these is the \textit{naive neighborhood} model: Let $r_p$ denote the distance between a particle and its nearest neighbor. For some tolerance $\tau \geq 0$, define the \textit{naive neighborhood} $\Ncal_\tau(p)$ of a particle $p$ as the set of all particles except $p$ whose distance to $p$ is less than or equal to $(\tau+1) r_p$. This model assumes that the immediate surrounding of a particle takes the form of a $D$-dimensional ball, that is, that the neighborship of every particle at a given distance from it is equally justified. As illustrated in \mbox{Fig. {\ref{fig:neighborhood_models}}}, however, this is not always the case, not even approximately.

A more natural neighborhood model can be derived from the Voronoi cell,\cite{devadoss_2011} variously called the Dirichlet region, Thiessen polytope, and Wigner-Seitz cell. Let $C(p)$ denote the Voronoi cell of a particle $p$. Then, its \textit{Voronoi neighborhood} $\Vcal(p)$ is defined as the set of all particles $q_i \neq p$ whose Voronoi cells $C(q_i)$ are adjacent to the Voronoi cell $C(p)$ of particle $p$.

Universal application of the Voronoi neighborhood model is precluded by its topological instability.\cite{tanemura_1977} A common approach to reducing this instability is to omit cell faces with a small area.\cite{sheng_2006, stukowski_2012, larsen_2016} Here, we describe a stochastic alternative, which we call the \textit{robustified} Voronoi neighborhood $\mathcal{V}^*$. Informally, it works by subjecting particles to Gaussian perturbations of some scale $\sigma \ll r_p$ and admitting to the neighborhood of a particle $p$ only those elements of its naive neighborhood $\Ncal_\tau(p)$ that appear in its post-perturbation Voronoi neighborhood 
with a probability of 50\% or more (see Appendix \ref{appx:formal_demarcation} for the formal definition). The choices of $\sigma$ and $\tau$ are discussed in \mbox{Sec. \ref{sec:computation}}.

We conclude this section by discussing one assumption that is implicit in our model of systems. In particular, by inferring neighborship in the way described, we assume that all prevailing interactions---whether ionic, covalent, metallic, or van der Waals---occur at the same length scale. In many cases, this assumption can be satisfied with suitable particle choice. When studying water, for instance, {\ce{H2O}} molecules, rather than {\ce{O}} and {\ce{H}} atoms, would be the suitable choice of particle, since the interactions between the latter constituents occur with different length scales at the inter- and intramolecular levels.

\section{Quantifying local orientational order}
\label{sec:quantifying_order}

{All spatial order arises from the breakdown of the spatial symmetries, and its physical origins can be energetic as well as entropic.{\cite{tanaka_2019}} \textit{Local} spatial order presents a genuine characterization challenge as, unlike the global variety, it is not conducive to traditional wave vector analysis. Particularly revealing of a material's structure is local \textit{orientational} order (LOO), which concerns the distribution of a particle's neighbors over its coordination shell. Given a reference structure, the quantification of LOO relative to that reference has at least one natural solution.{\cite{martelli_2018}} However, the articulation of LOO as an \textit{absolute} quantity has seen little progress since the seminal work of Steinhardt and coworkers {\cite{steinhardt_1983}} almost 40 years ago. Below, we address this challenge with an approach inspired by the observation of a certain kind of informational redundancy in the geometry of regular polyhedral neighborhoods.}

\subsection{Bond angles}
\label{sec:bond_angles}

A complete description of a particle's neighborhood can be given by the set of vectors expressing the relative position of its neighbors. Given a particle $p$ and a neighboring particle $q$, call the vector $q - p$ a \textit{bond}. Notice that rotations and reflections preserve the geometric form of a neighborhood but not the bonds that describe it. It is thereby evident that the geometric form of a particle neighborhood is encoded in the \textit{relationships between bonds} rather than the \textit{bonds} themselves.

In the present work, we capture the relationships between bonds through the \textit{bond angle}, defined as the smaller angle between two bonds. 

Observe that neighborhoods with the form of a regular convex polytope exhibit a distinct lack of diversity in bond angles. Take the icosahedral neighborhood, which is described by $12$ bonds: the $66$ pairs made by these bonds meet at one of only three different angles, namely those of (approximately) $63.4$\textdegree, $116.6$\textdegree, and $180$\textdegree. These bond pairs are highly redundant in the sense that only three different angles are needed to describe all $66$ of them. While this particular redundancy is readily attributed to the $60$ rotational symmetries of the icosahedron, a natural question to ask is whether the extent of such a redundancy conveys the absolute degree of orientational order around a particle in general. We here hypothesize in the affirmative and call this kind of redundancy \mbox{\textit{extracopularity}}.

\subsection{Extracopularity coefficient}
\label{sec:extracopularity}

Extracopularity can be quantified as the amount of information one saves by expressing the geometry of a particle neighborhood through bond angles rather than bond pairs. We now define such a quantifier of extracopularity and examine some of its properties.

\subsubsection{Definition}
\label{sec:definition}

A fundamental idea from information theory \cite{cover_2012} is that the \textit{information content} of a set $A$ corresponds to the number of binary (or, more generally, $d$-ary) variables needed to enumerate its elements.\cite{hartley_1928, shannon_1948} This number is given by the \textit{Hartley information} $I(A) = \log_2 \abs{A}$ bits, where $\abs{A}$ denotes the \textit{cardinality} or number of elements in $A$. Let $p$ be a particle. Define its \textit{coordination number} ${k}$ as the number of bonds that it forms, and let $B$ denote its set of (unordered) bond pairs. From elementary combinatorics, we know that, without replacement, one can make $(k^2 - k)/2$ pairs from $k$ elements. The information content of bond pairs for $p$ is therefore
\begin{equation}
    I(B) = \log_2 \left[\frac{{k^2 - k}}{2}\right].
\end{equation}
Let $\Theta$ denote the set of all bond angles for particle $p$ and $m = \abs{\Theta}$ its cardinality. Then, its information content is simply
\begin{equation}
    I(\Theta)= \log_2 (m).
\end{equation}
Now, the amount of information one saves by accounting for bond angles rather than bond pairs can be written as a difference as follows:
\begin{equation}
    E := I(B) - I(\Theta).
\end{equation}
We call this difference the \textit{extracopularity coefficient} of the particle $p$, given in simpler form by
\begin{equation}
    E = \log_2  \left[\frac{ k^2 - k}{ 2 m} \right], \quad k > 1.
\end{equation}

The extracopularity coefficient corresponds to the conditional Hartley information of bond pairs given bond angles. This quantifies how much easier the hypothetical binary search for a particular bond pair would, on average, be if the angle made by the pair were known. A large $E$ thus tells us that a neighborhood is simple in the sense of lacking diversity in bond angles. Based on its information-theoretic interpretation, it makes conceptual sense to define $E = 0$ for $k=1$, which we hereafter do.

To simplify the exposition, we have thus far taken the bond angle to be a discrete quantity while it is, in fact, a continuous one. Due to heat, a practical system will almost never exhibit two bond pairs with precisely the same angles. Hence, strictly speaking, for every particle of any practical system, we have trivially
\begin{equation}
    m = (k^2-k)/2 \quad \Rightarrow \quad E = 0.
\end{equation}
Clearly, any redundancy that exists in the geometry of a particle's neighborhood will be hidden under a layer of thermal noise. One can expose such redundancies through the discretization of bond angles. This, however, is a difficult problem to solve exactly, owing to the notorious intransitivity of similarity relations.\cite{poincare_1905, bandelt_1992} Fortunately, there is a fast heuristic alternative which we discuss in \mbox{Sec. \ref{sec:number_of_bond_angles}}.

\subsubsection{Properties}
\label{sec:properties}

Let us examine a few mathematical properties of $E$. First consider its behavior with respect to the coordination number $k$. For a given $m$, it is readily seen that $E$ increases monotonically with respect to $k$. The coefficient is thus guaranteed to be able to indicate the boundaries of an otherwise perfect crystal, as particles in such regions differ from their interior counterparts only by $k$. In the general case where $m$ is not fixed, however, the relationship between $E$ and $k$ is unclear. Nevertheless, there are important special cases where this relationship can be precisely obtained. 

Consider, for instance, neighborhoods of regular convex polygonal form, where we have $m=k/2$ if $k$ is even and $m=(k-1)/2$ if $k$ is odd. Visibly,
\begin{equation}
    E_{\, \tsub{regular \\ convex \\ polygonal}} =
    \begin{cases}
        \log_2 \left( k -1 \right) &  \text{if } k \text{ is even}, \\
        \log_2  \left(k  \right)    & \text{if } k \text{ is odd}. \\
    \end{cases}
\end{equation}
Two intuitive properties of $E$ follow immediately from this result: (I) it increases monotonically with the degree of rotational symmetry in regular convex polygonal neighborhoods, and (II) these increases decay like the reciprocal of the degree of rotational symmetry in such neighborhoods. 

\begin{table*}
    \centering
    \begin{ruledtabular}
    \begin{tabular}{lllllrrl} \multicolumn{3}{c}{\centered{Description}} & \multicolumn{2}{c}{Corresponding structures} &  \multicolumn{3}{c}{Parameters} \\ \cmidrule{1-3} 
    \cmidrule{4-5} \cmidrule{6-8} Abbrev. & Geometry & Classification & Lattice & TCC\cite{malins_2013_b} & \multicolumn{1}{l}{$k$} & \multicolumn{1}{l}{$m$} & $E$ \\\midrule
    TBP & Trigonal bipyramidal & Deltahedral, bipyramidal & Honeycomb &  & $5$ & $3$ & $1.737$ \\
    SDS & Snub disphenoidal\footnote{Often called trigonal dodecahedral} & Deltahedral & & & $8$ & $6$ & $2.222$ \\
    PBP & Pentagonal bipyramidal & Deltahedral, bipyramidal &  & & $7$ & $4$ & $2.392$ \\
    CTP & Capped trigonal prismatic & Prismatic &  & & $7$ & $4$ & $2.392$ \\
    BTP & Bicapped trigonal prismatic & Prismatic & & & $8$ & $5$ & $2.485$ \\
    TET & Regular tetrahedral & Platonic, deltahedral & Diamond cubic & & $4$  & $1$ & $2.585$  \\
    HBP & Hexagonal bipyramidal & Bipyramidal & Simple hexagonal & & $8$ & $4$ & $2.807$ \\ 
    CSA & Capped square antiprismatic & Antiprismatic & & & $9$ & $5$ & $2.848$ \\
    CSP & Capped square prismatic & Prismatic & & & $9$ & $5$ & $2.848$ \\
    TTP & Tricapped trigonal prismatic & Prismatic, deltahedral & & & $9$ & $5$ & $2.848$ \\
    SC & Regular octahedral\footnote{Square bipyramidal} & Platonic, deltahedral, bipyramidal & Simple cubic &  & $6$ & $2$ & $2.907$  \\
    BSA & Bicapped square antiprismatic & Deltahedral, antiprismatic & & 11A & $10$ & $6$ & $2.907$ \\
    BSP & Bicapped square prismatic & Prismatic & & & $10$ & $5$ & $3.170$ \\
    CPP & Capped pentagonal prismatic & Prismatic & & & $11$ & $6$ & $3.196$ \\
    SA & Square antiprismatic & Antiprismatic & & & $8$ & $3$ & $3.222$   \\
    HXD & Regular hexahedral\footnote{Square prismatic, cubic} & Platonic, prismatic & Body-centered cubic & 9X & $8$ & $3$ & $3.222$ \\
    BPP & Bicapped pentagonal prismatic & Prismatic & & 13B & $12$ & $7$ & $3.237$ \\
    HCP & Anticuboctahedral\footnote{Triangular orthobicupolar} & Bicupolar & Hexagonal close-packed  & HCP & $12$ & $6$ & $3.459$ \\
    BCC & Rhombic dodecahedral & Catalan & Body-centered cubic & BCC & $14$ & $6$ & $3.923$  \\
    FCC & Cuboctahedral\footnote{Triangular gyrobicupolar} & Bicupolar & Face-centered cubic & FCC & $12$ & $4$ & $4.044$ \\
    CPA & Capped pentagonal antiprismatic & Antiprismatic & & 12B & $11$ & $3$ & $4.196$  \\
    ICO & Regular icosahedral\footnote{Bicapped pentagonal antiprismatic} & Platonic, deltahedral, antiprismatic & & 13A & $12$ & $3$ & $4.459$ \\
    \end{tabular}
    \end{ruledtabular}
    \caption{Commonly encountered geometries of coordination, listed in order of increasing $E$, rounded to three decimal places. These comprise the first $4$ Platonic solids, all $8$ convex deltahedra; $12$ capped (anti)prisms, $4$ regular bipyramids, $2$ circumscribable bicupolae, and the rhombic dodecahedron (a Catalan solid). Where possible, corresponding structures are provided. Of the two geometries that correspond to the BCC lattice, in our approach, only the rhombic dodecahedral is interpreted as such.
    }
    \label{table:common_geometries}
\end{table*}

While $E$ does not have a (known) closed-form solution in terms of $k$ for neighborhoods of regular convex poly\textit{hedral} (i.e. Platonic) form, given their small number, we can study such neighborhoods by exhaustion. The coefficients of the first four Platonic geometries are given in Table \ref{table:common_geometries}. {We conjecture that the fourth one (icosahedral) maximizes $E$ under the constraint of equal sphere packing (i.e. $r_q \approx r_p$ for all neighbors $q$ of $p$)}. The fifth and final one (dodecahedral), which does not satisfy this constraint, has $E \approx 5.248$. Visibly, $E$ is increasingly monotonic in $k$ also for Platonic neighborhoods. Furthermore, a (weaker) ordinal association between $E$ and $k$ is observed over all geometries in Table {\ref{table:common_geometries}}. In particular, nearly two-thirds of the former's variation can be explained by the latter.

Finally, consider the geometric form of neighborhoods that maximize $E$ for a given $k$. It is readily seen that the inequality
\begin{equation}
    E \leq  \log_2 \left[\frac{ k^2 - k}{ 2} \right]
\end{equation}
is satisfied with equality if and only if there is a single unique bond angle. It can be shown this is true only when the neighborhood has the form of a regular simplex, corresponding to an equilateral triangle in 2D and a regular tetrahedron in 3D. That $E$ is maximized by such geometries is a rather natural property given that regular simplices constitute the simplest possible polytopes.

\section{Computation}
\label{sec:computation}

We have developed an algorithm for computing extracopularity coefficients for three-dimensional systems, a prototype implementation of which is publicly available.\footnote{Code available at \url{www.github.com/johncamkiran/extracopularity}} On a consumer-grade computer, this prototype is able to process a system of one million particles in a minute. Below, we discuss three nontrivial aspects of our algorithm.

\subsection{Calculating membership probabilities}
\label{sec:membership_probabilities}

Recall that determining a robustified Voronoi neighborhood requires the probability that a particle belongs to an (ordinary) Voronoi neighborhood after receiving a Gaussian perturbation of scale $\sigma$. This probability is straightforward to evaluate in the Monte Carlo way, that is, by sampling the perturbations (see Appendix \ref{appx:membership_prob} for details). The number of samplings $M$ needed for the probability to converge depends on the geometries of the underlying neighborhoods. The results in \mbox{Sec. \ref{sec:neighborhood_demarcation}} suggest $M=4$ to be a suitable universal choice. For redundancy, the default choice of our algorithm is $M=8$. As for $\sigma$, we set ${\sigma =\avg{r_p} /5}$ on account of its apparent optimality for neighborhood demarcation. We note, however, that other values around this choice of $\sigma$ work just as well. 

\subsection{Determining the number of different bond angles}
\label{sec:number_of_bond_angles}

A fast and accurate method for determining the bond angle count $m$ is to compare the observed bond angles of a particle to those of a commonly encountered geometry (CEG), such as the ones in \mbox{Table \ref{table:common_geometries}}. If a match is found, $m$ can be set directly to its known value for the commonly encountered geometry (CEG). To achieve this, we compute, for every CEG, the root-mean-square error (RMSE) between the observed bond angles of the particle and those of the CEG. If for any CEG, the RMSE lies below a cutoff level, we assign the particle the bond angle count of the CEG that produces the smallest RMSE (see Appendix \ref{appx:neighborhood_error} for details). The cutoff level is chosen to reproduce the FCC volume fraction indicated by polyhedral template matching performed with a threshold of $0.15$.

At little computational cost, this approach can account for a large yet finite number of geometries. A precise method for dealing with unrecognized geometries remains to be found. However, since $k$ appears to to be able to explain an important part of the variation in $E$, an idea of $E$ can be obtained with an estimator $\hat{m}$ of the number of bond angles given $k$. Our algorithm uses the following one (see Appendix {\ref{appx:bond_angle_count_estimate}} for its derivation):
\begin{equation}
    \hat{m}(k) \approx 7.3 - 11.4\exp(-k/3.4) 
\end{equation}
This estimator is devised to give a conditional lower bound on the bond angle count given that no CEGs are detected.

\subsection{Restriction to the naive neighborhood}
\label{sec:naive_neighborhood_restriction}

Recall that the robustified Voronoi neighborhood $\mathcal{V}^*(p)$ is restricted to the naive nearest neighborhood $\mathcal{N}_\tau(p)$, which has a single parameter $\tau$ controlling its tolerance. Our algorithm sets $\tau$ based on the location of the first minimum in the radial distribution function of a Lennard-Jones crystal at nominal temperature and pressure.\cite{tenWolde_1995, malins_2013_b} This corresponds to $\tau = 0.50$ for BCC and $\tau = 0.36$ otherwise. {One issue with this choice is that it gives ${\langle k \rangle = 12}$ for random close-packed (RCP) systems, which we know could not be true. To resolve this issue we take ${\tau = 0.20}$ for neighborhoods with unrecognized geometries, which is approximately the choice for which the fraction of $12$-coordinate particles in RCP systems equals the number of $12$-coordinate CEGs detected. Our piecewise choice of $\tau$ can be summarized as follows:}
\begin{equation}
\tau := 
    \begin{cases}
    0.50 & \text{if BCC angles are observed}, \\
    0.36 & \text{if other CEG angles are observed}, \\
    0.20 & \text{otherwise}.
    \end{cases}
\end{equation}

\section{Validation}
\label{sec:validation}
In order to validate our algorithm, we considered two aspects of its performance for which ground truth is available.

\subsection{Accuracy in demarcating lattice neighborhoods}
\label{sec:neighborhood_demarcation}

\vspace{-0.5\baselineskip}

We started by testing the convergence of our Monte Carlo method for determining robustified Voronoi neighborhoods. We performed $1000$ trials of this test for common lattice types; \mbox{Table \ref{table:samplings_needed}} summarizes the results. For each lattice, rapid convergence to the analytical Voronoi neighborhood was observed.

\begin{table}[!b]
    \centering
    \begin{ruledtabular}
    \begin{tabular}{llrr}
         Abbreviation & Lattice & $M$ & $k$ \\ \midrule
         BCC & Body-centered cubic & $3$ & $14$ \\
         FCC & Face-centered cubic & 3 & 12 \\
         HCP & Hexagonal close-packed\footnote{$a_3/a_1 = \sqrt{8/3}$ \ (ideal).}  & $3$ & $12$ \\
         SC & Simple cubic & $2$ & $6$ \\
         PM & Primitive monoclinic\footnote{$a_2/a_1 = 4/3$, \ $a_3/a_1 = 3/2$, \ and  $\beta = 50^{\circ}$.} & $2$ & $6$ \\
         DC & Diamond cubic & $1$ & $4$
    \end{tabular}
    \end{ruledtabular}
    \caption{Number of samplings $M$ needed for the Monte Carlo solution of our neighborhood model to converge and the coordination number $k$ upon convergence.
    }
    \label{table:samplings_needed}
\end{table}

\begin{table}[!b]
    \centering
    \begin{ruledtabular}
        \begin{tabular}{lrrrrrr}
            \multirow{2}{*}{$T/T_\text{m}$}  & \multicolumn{3}{c}{Probability (\%)} & \multicolumn{3}{c}{RMAD (\%) } \cr\cmidrule{2-4} \cmidrule{5-7}
                      & FCC    & HCP    & BCC    & FCC   & HCP    & BCC  \cr \midrule
                $0.2$ &  $0.0$ &  $0.0$ & $0.0$  & $0.0$ &  $0.0$ &  $0.0$ \cr
                $0.4$ &  $0.1$ &  $1.5$ & $1.8$  & $0.0$ &  $0.2$ &  $0.6$ \cr
                $0.6$ &  $3.3$ & $12.2$ & $13.1$ & $1.0$ &  $1.8$ &  $4.6$ \cr
                $0.8$ & $20.0$ & $37.6$ & $34.3$ & $6.6$ & $7.0$ & $12.3$
        \end{tabular}
    \end{ruledtabular}
\caption{Probability of a deviation and the relative mean absolute deviation (RMAD) in $E$ for three crystals at various fractions of their melting point $T_\text{m}$.}
\label{table:deviation_statistics}
\end{table}

\begin{figure*}[!t]
    \centering
    \begin{tabular}{lc@{\hskip 0.33cm}c@{\hskip 0.33cm}c}
        & $E$ & $Q_6$ & PTM \\ \\
        \centered{\multirow{1}{*}{(a)}} & \centered{\includegraphics[width=0.28\linewidth]{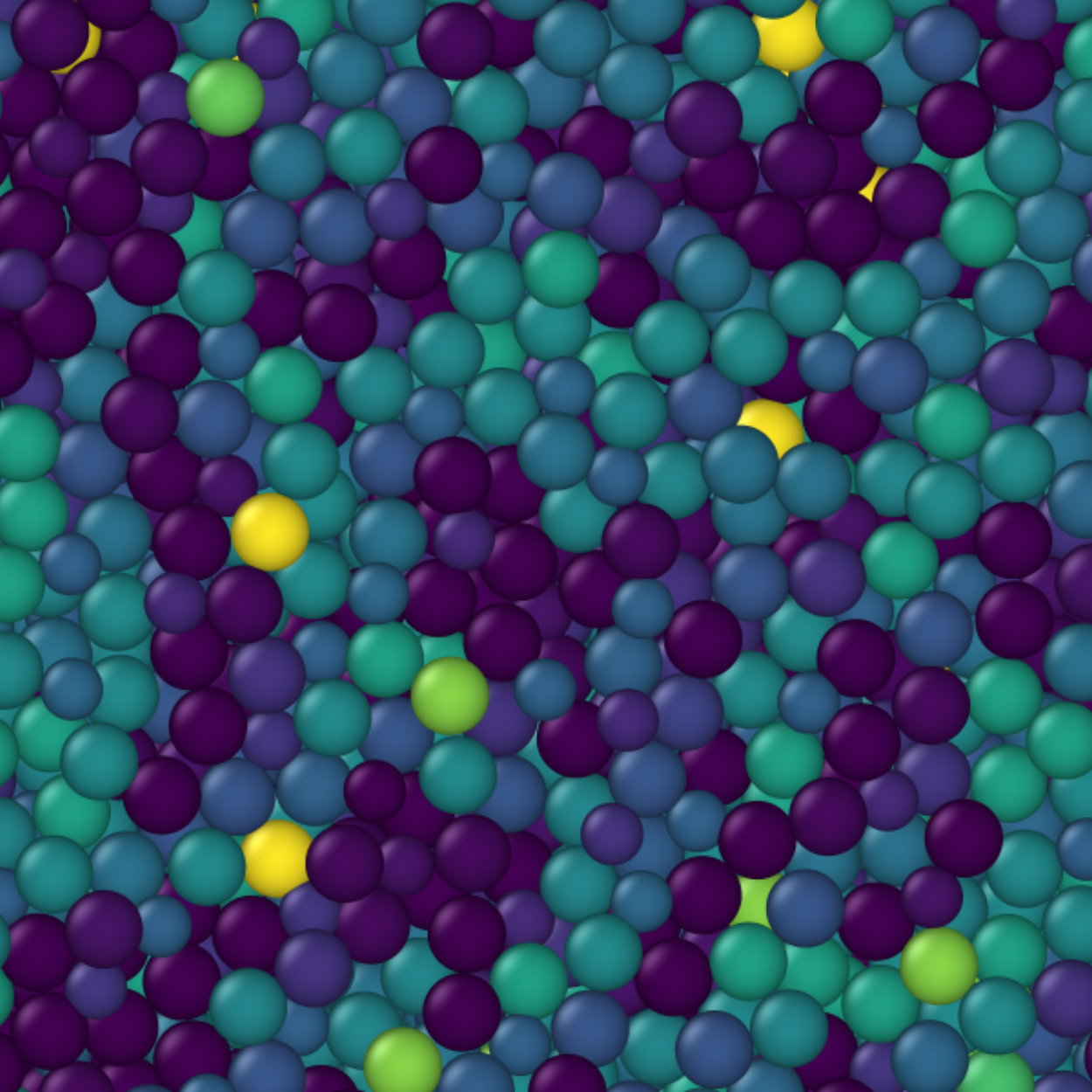}} & \centered{\includegraphics[width=0.28\linewidth]{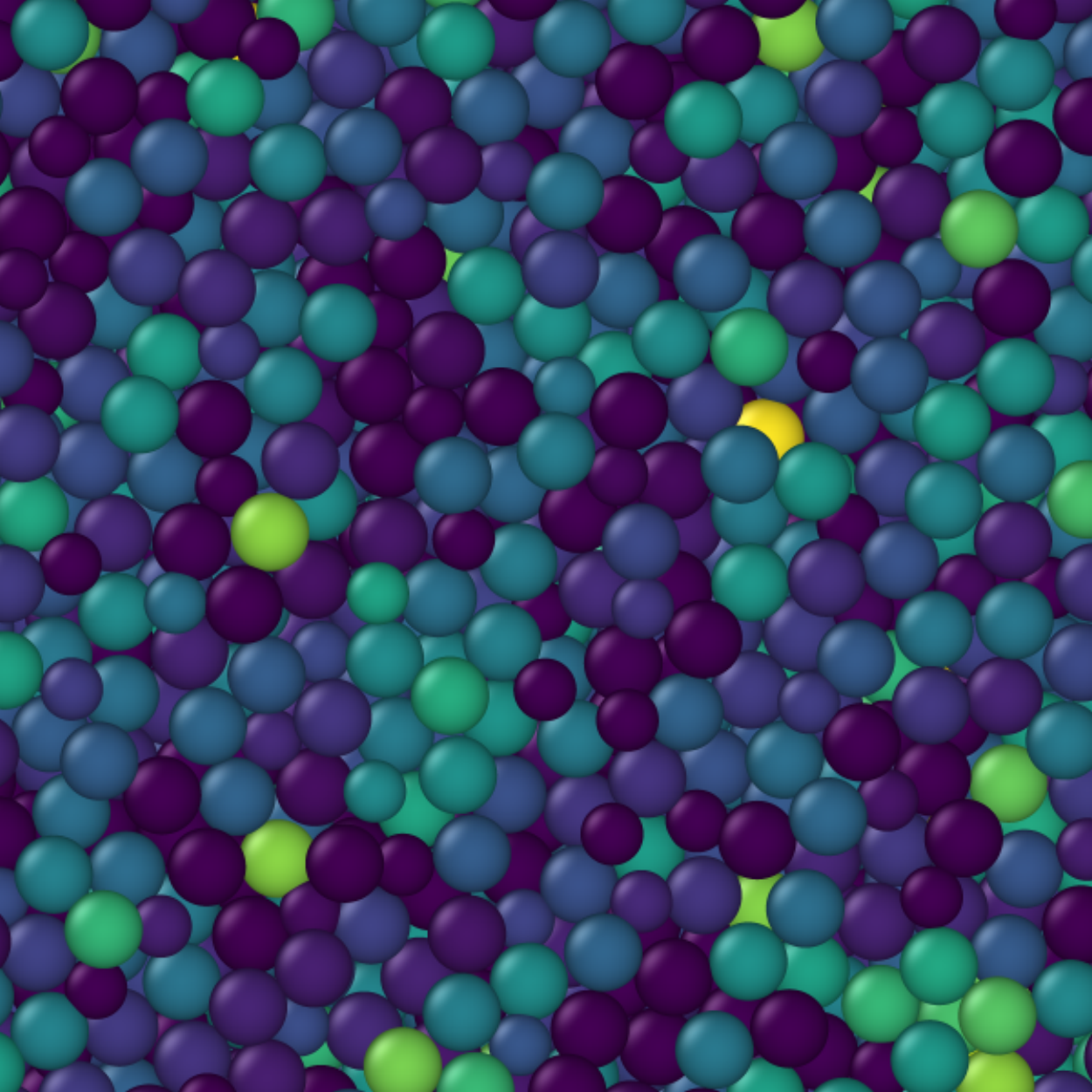}} &
        \centered{\includegraphics[width=0.28\linewidth]{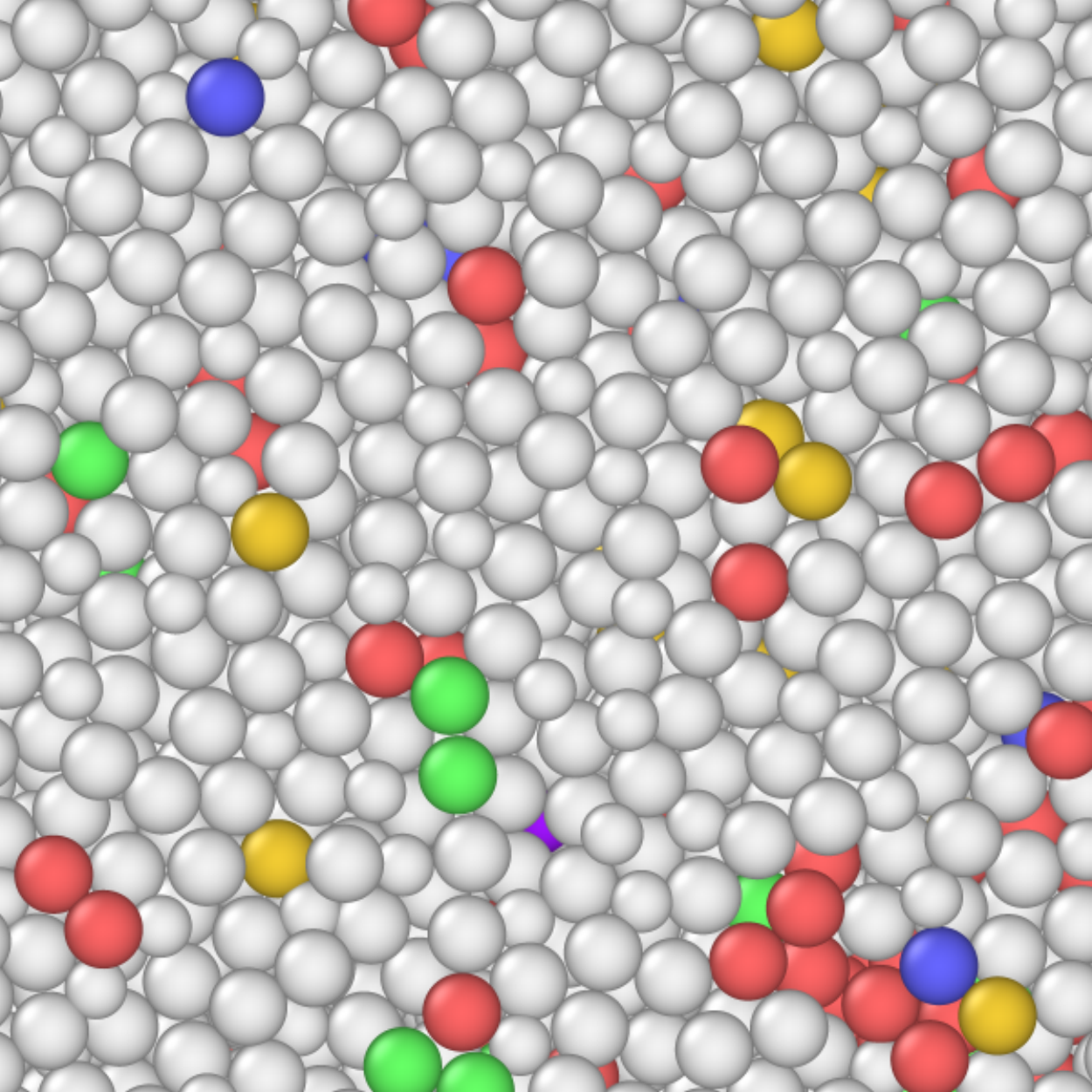}}\\ \\ 
        \centered{\multirow{1}{*}{(b)}} & \centered{\includegraphics[width=0.28\linewidth]{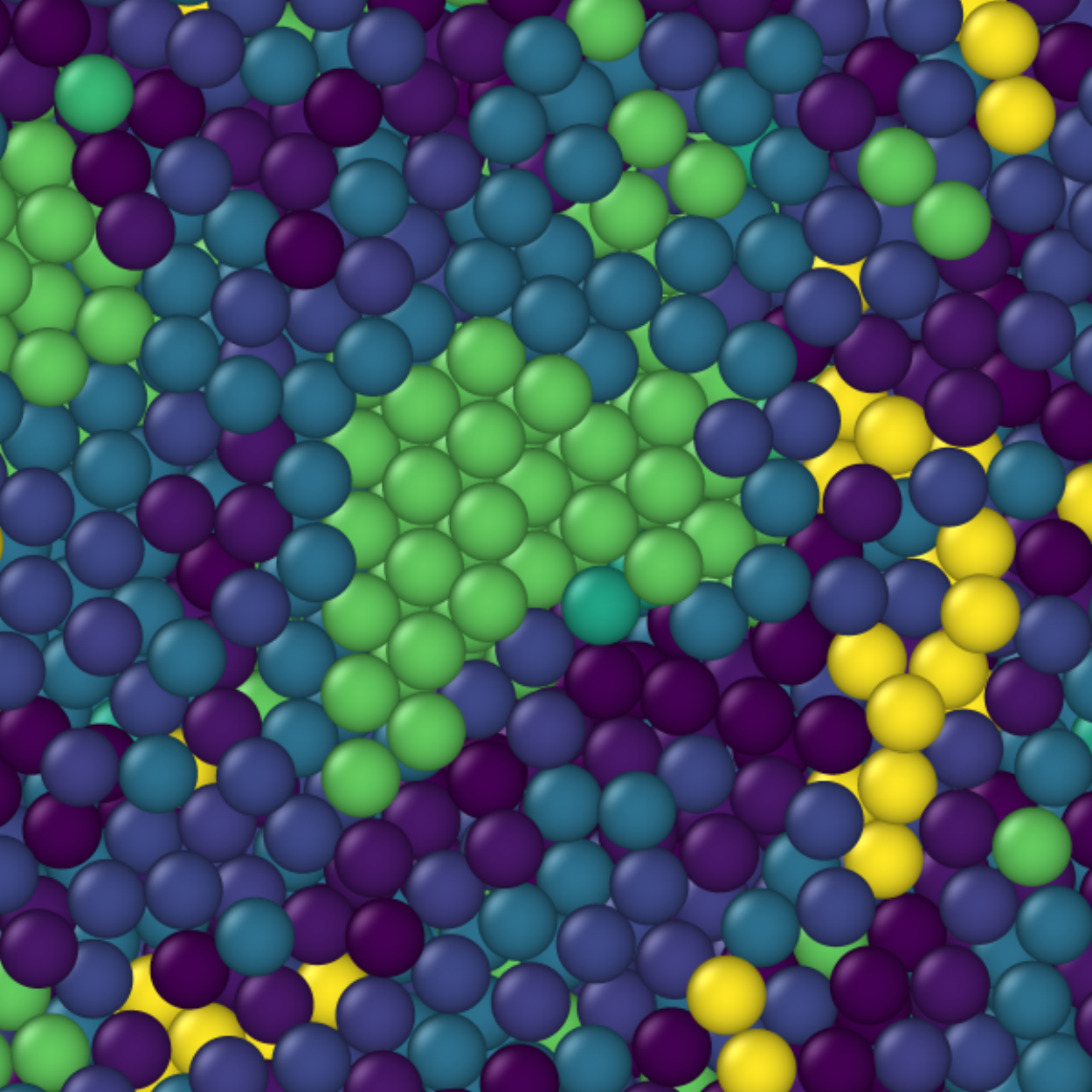}} & \centered{\includegraphics[width=0.28\linewidth]{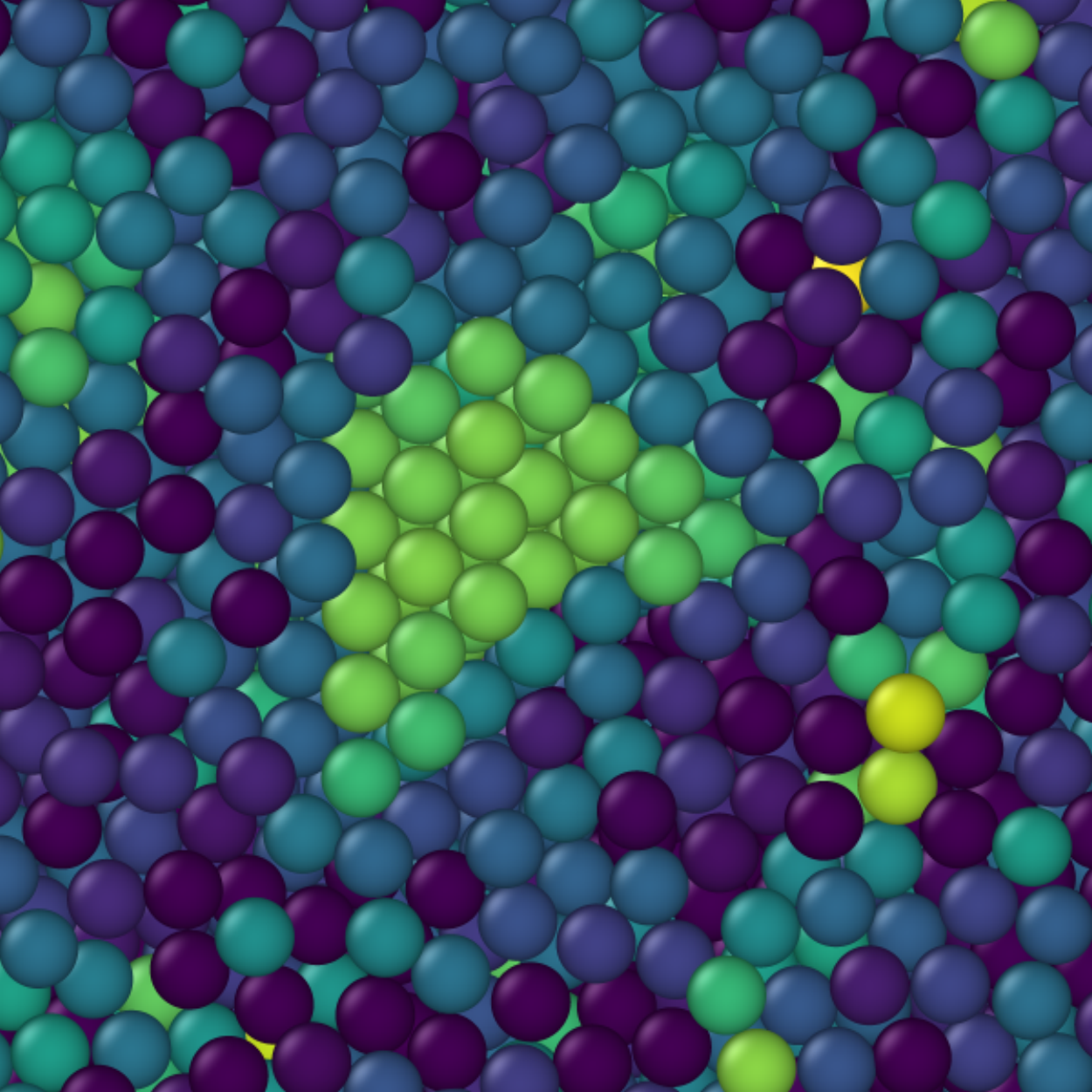}} &
        \centered{\includegraphics[width=0.28\linewidth]{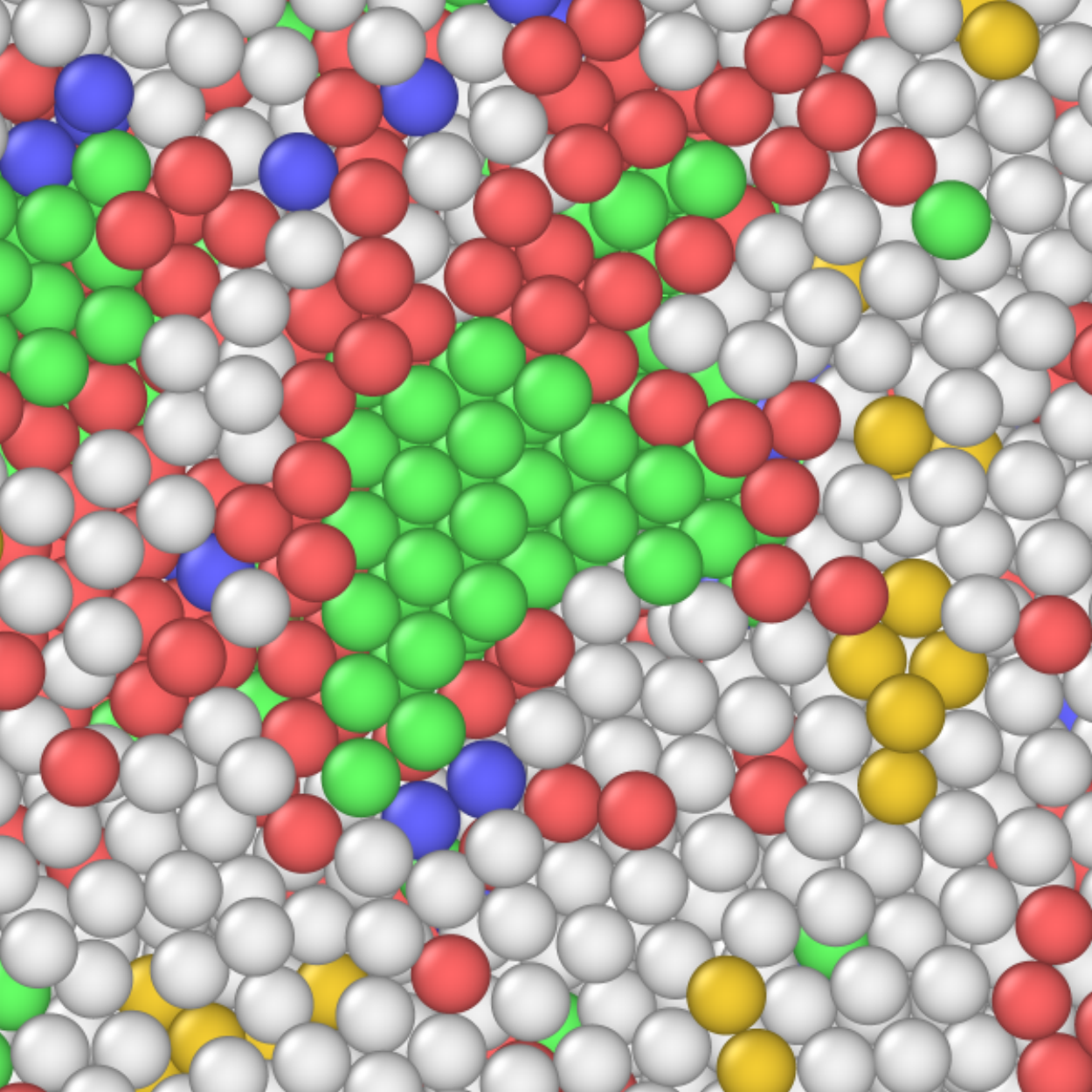}} \\
        \\ 
        \centered{\multirow{1}{*}{(c)}} & \centered{\includegraphics[width=0.28\linewidth]{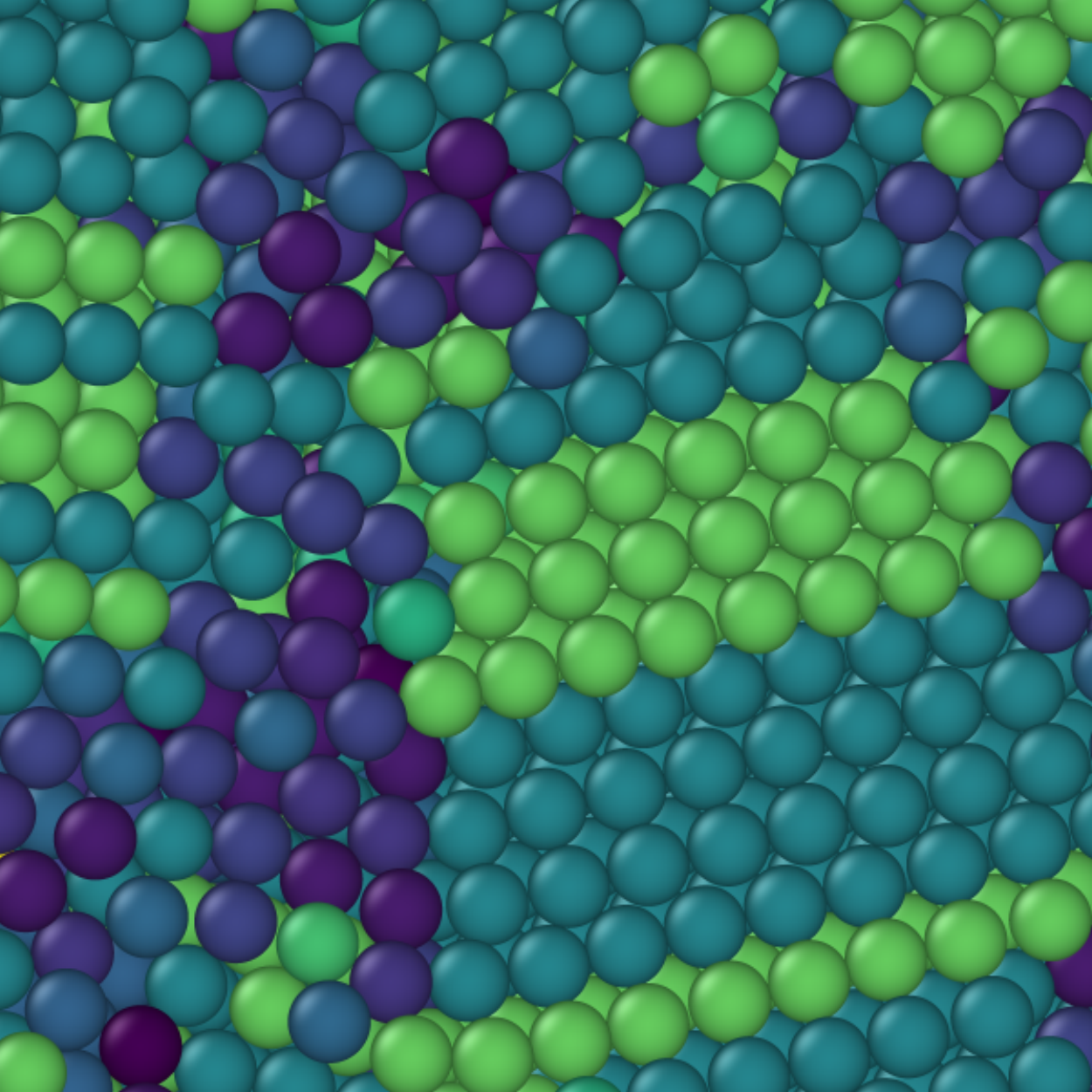}} & \centered{\includegraphics[width=0.28\linewidth]{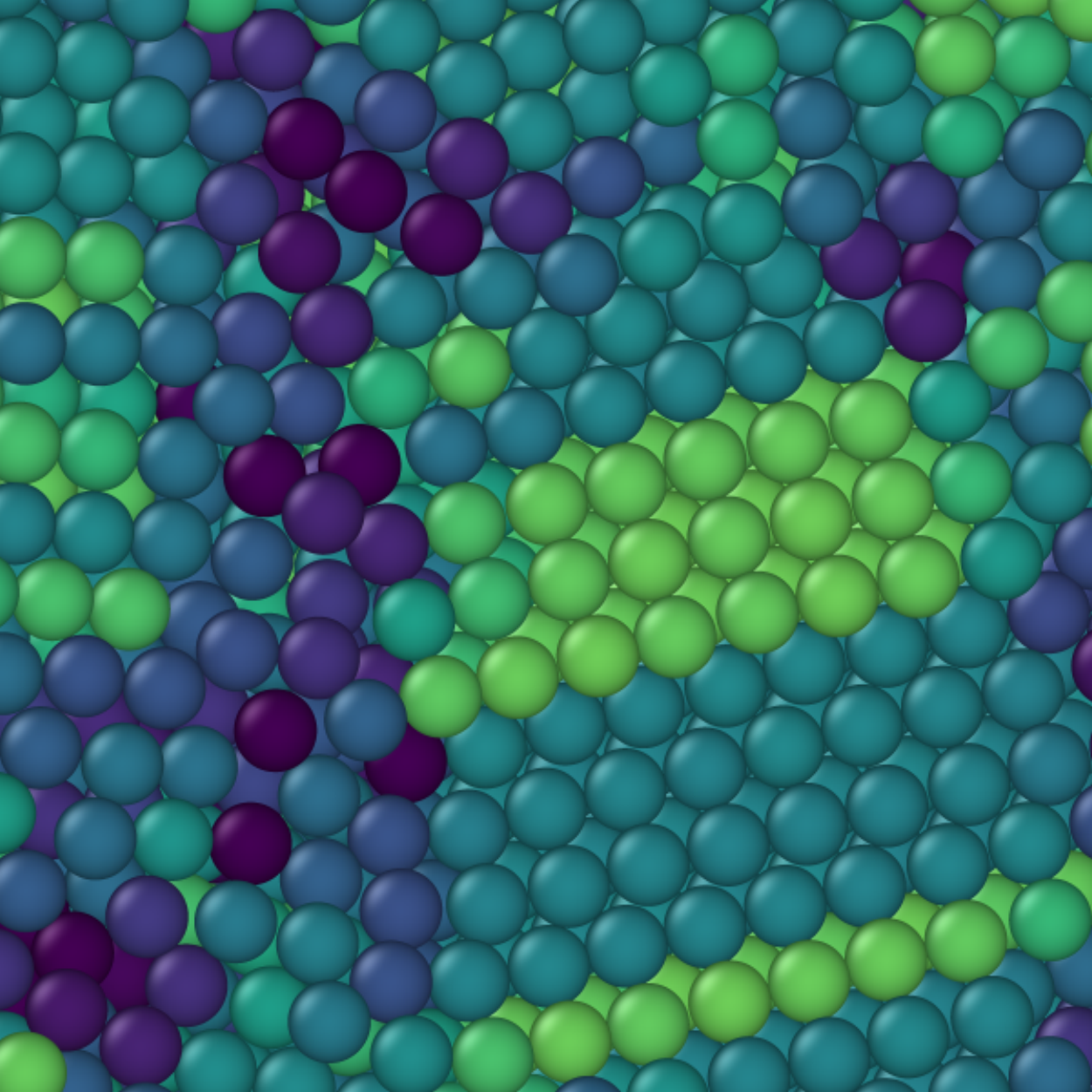}} &
        \centered{\includegraphics[width=0.28\linewidth]{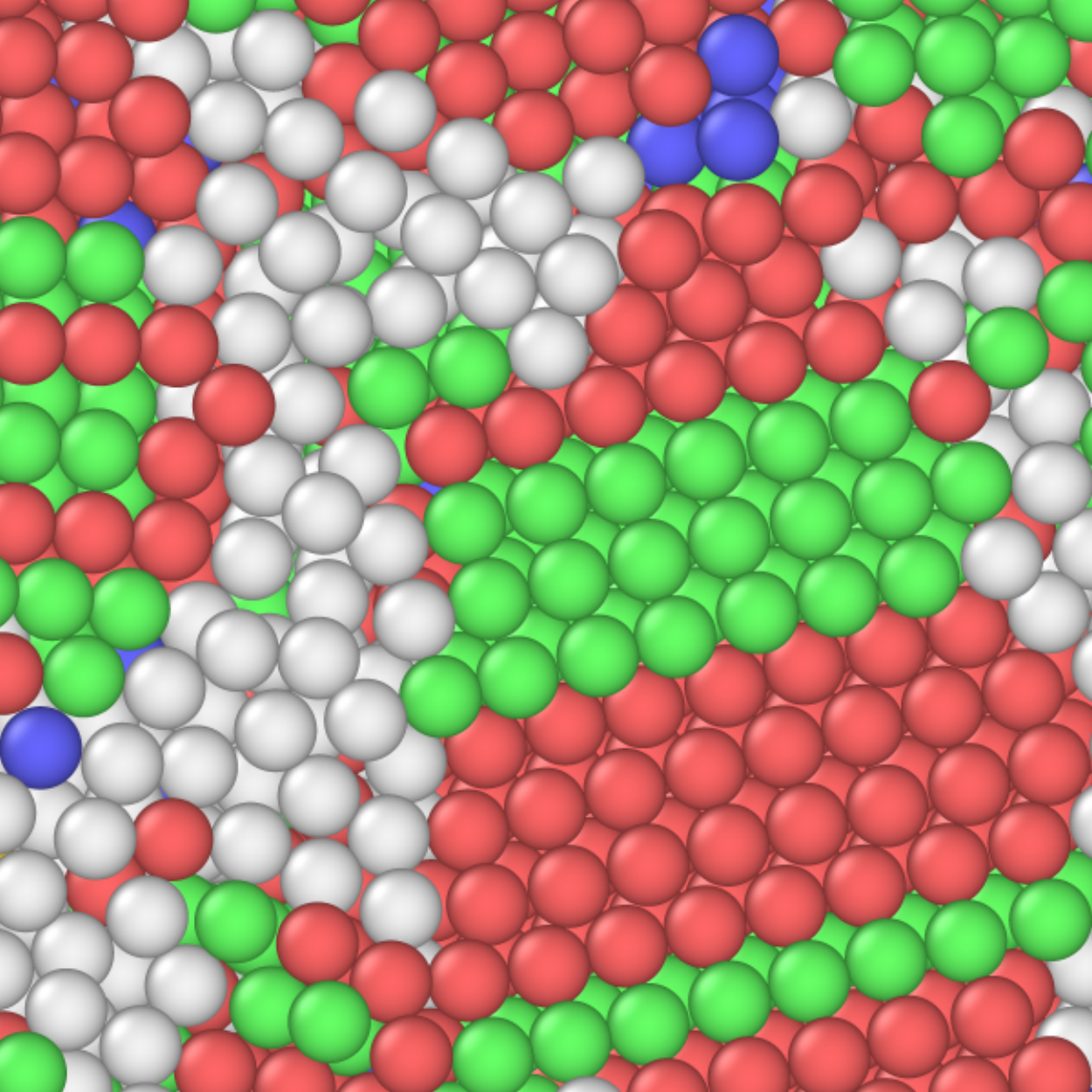}}
    \end{tabular}
    \caption{Particles colored by $E$, $Q_6$, and PTM for the (a) KA glass, (b) LJ glass, and (c) LJ crystal. For $E$ and $Q_6$, lighter colors imply larger values. The upper bound of their color map (yellow) corresponds to icosahedral order. The lower bound (indigo) has been chosen to produce similar contrast in order to facilitate comparison. In PTM, green indicates FCC, red indicates HCP, blue indicates BCC, yellow indicates ICO, and violet indicates SC. PTM was performed with a threshold of $0.20$ in (a) and $0.15$ in (b) and (c). All images were rendered in OVITO.\cite{stukowski_2009}}
    \label{fig:particles_colored_by_E}
    \makeatletter
\let\save@currentlabel\@currentlabel
\edef\@currentlabel{\save@currentlabel(a)}\label{fig:particles_colored_by_E_a}
\edef\@currentlabel{\save@currentlabel(b)}\label{fig:particles_colored_by_E_b}
\edef\@currentlabel{\save@currentlabel(c)}\label{fig:particles_colored_by_E_c}
\makeatother
\end{figure*}

\subsection{Robustness to thermal fluctuations}
\label{sec:thermal_performance}

\vspace{-0.5\baselineskip}

Next we tested the robustness of our algorithm to heat. Local structural indicators are often practically limited by their sensitivity to thermal fluctuations. Hence, a common practice is to briefly quench {\cite{stillinger_1983}} a system before analyzing it. We tested the performance of our algorithm without the use of such techniques so as to obtain a lower bound on its performance. We simulated FCC, HCP, and BCC crystals in equilibrium at various temperatures. The former two were simulated using the Lennard-Jones potential, while the latter was simulated through an embedded-atom potential.\cite{mendelev_2003} \mbox{Table \ref{table:deviation_statistics}} provides a statistical summary of deviations in $E$ from its value at absolute zero temperature. The indicator was found to be unaffected by low temperatures and fairly accurate up to intermediate temperatures. 

\section{Application}
\label{sec:application}

{Having validated our algorithm, we subsequently applied it to the analysis of three systems whose exact structure is not known \textit{a priori}. We compared our results to the Steinhardt order parameter{\cite{steinhardt_1983}} $Q_6$ and polyhedral template matching{\cite{larsen_2016}} (PTM), a widely used structural classifier for crystalline materials. Particle position data for the three systems were obtained through molecular dynamics simulations performed with LAMMPS.{\cite{plimpton_1995}} Each simulation comprised $125,000$ particles in a cubic box with periodic boundary conditions. Below, all physical quantities are stated in reduced units.

For the first system, we simulated the liquid--glass transition of a Kob-Andersen (KA) binary mixture of Lennard-Jones fluids.{\cite{kob_1995,pedersen_2018}} We initialized the system in an isothermal--isobaric (NPT) ensemble with a temperature of $T_1 = 2$ and a pressure of $P = 10.19$. The temperature was then lowered under constant pressure to $T_2 = 0.1$ at a rate of $\Delta T / \Delta t = 0.019$. The transition was observed to occur at $T_g \approx 0.6$. For the second system, we repeated this simulation with a single-component Lennard-Jones (LJ) fluid. For the third system, we simulated the crystallization of the single-component LJ liquid, achieved by repeating the second simulation with a cooling rate that is $10$ times slower. The results of these simulations are depicted in Fig. {\ref{fig:particles_colored_by_E}}.

In the KA glass, $E$ indicated a notable concentration of bicapped square antiprismatic neighborhoods, which are known to be locally favored in KA mixtures. {\cite{malins_2013_a}} Interestingly, however, (mono)capped square antiprismatic neighborhoods were detected in roughly equal quantity. These ordered regions were separated by networks of relative disorder, depicted by the darkest particles in Fig. {\ref{fig:particles_colored_by_E_a}}. The results of $E$ showed some agreement with $Q_6$, especially over the minority of particles with ICO neighborhoods. PTM showed no evidence of crystallization, consistent with the tendency of KA mixtures.

In the LJ glass, $E$ detected several small, predominantly FCC crystallites, one of which is depicted in {\ref{fig:particles_colored_by_E_b}}. It also indicated a significant incidence of the ICO geometry, which is known to be locally favored in single-component LJ systems. The results of $E$ were comparable to those of $Q_6$ and PTM, except in distinguishing icosahedrality from crystallinity, where $Q_6$ often struggled.

The LJ crystal was found to be dense in planar defects, notably grain boundaries, stacking faults, and FCC--HCP interfaces. All three of these features are seen in Fig. {\ref{fig:particles_colored_by_E_c}}. On the latter two defects, $E$ showed strong agreement with PTM. Meanwhile its characterization of grain boundaries, which cannot be studied by PTM, was similar to that of $Q_6$. }

\section{Discussion}
\label{sec:discussion}

\begin{figure}[!b]
    \includegraphics[width=\linewidth]{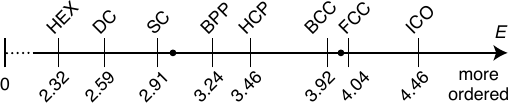}
    \caption{A few important geometries ranked by $E$. Here, HEX denotes the hexagonal geometry (2D).}
    \label{fig:extracopularity_axis}
\end{figure}

This work presents a local structural indicator called $E$ that quantifies a pairwise informational redundancy among the bonds formed by a particle. In doing so, it
appears to be able to rank particle neighborhoods based on their absolute degree of orientational order. This order ranking (depicted in \mbox{Fig. \ref{fig:extracopularity_axis}}) passes two important tests of soundness: (I) a given Bravais lattice is only ranked above another if it exceeds the other in point symmetry or packing efficiency, and (II) every Bravais lattice in $D = 3$ dimensions is ranked above all such lattices in $D = 2$ dimensions. Thus, $E$ exhibits basic agreement with crystallography. The ranking also seems to be consistent with the one implied by $Q_6$ as well as various empirical observations: the scarcity of crystals with the SC structure, the often unexplained preference of crystalline systems for the FCC over the HCP structure,\cite{heitkam_2012} and the minimum energy nature of the ICO geometry.\cite{frank_1952}

Our analytical results demonstrate that $E$ is in principle able to distinguish a wide range of geometries, and our computational work has given some evidence of its practicability. The main practical challenge is in determining the number of bond angles. For known geometries, our method is fast and accurate; for unrecognized geometries, it is at least able to give an effective lower bound.

\begin{figure}[!t]
    \centering
    \includegraphics{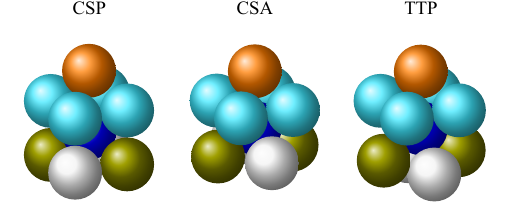}
    \caption{Three $9$-coordinate geometries with the same $E$.}
    \label{fig:conjugate_geometries}
\end{figure}

There are certain cases of distinct geometries being equal in $E$, as seen in Table {\ref{table:common_geometries}}. Caution must therefore be exercised in its interpretation. Such equalities in $E$, however, may not be entirely unjustified. They may indeed indicate an underlying geometric commonality as in the case of the CSP, CSA, and TTP (depicted in Fig. {\ref{fig:conjugate_geometries}}), the former two of which are equal up to a twist of base and the latter two of which possess a close and well-known resemblance. Where this is true, equalities in $E$ may be better thought of as conjugacies rather than degeneracies.

One limitation that is common to purely orientational indicators of local structure like $E$ is their inability to capture structure beyond the neighborhood boundary. Such indicators are thus unable to account for the translational aspect of local order, which appears crucial for certain systems, such as tetrahedral fluids.\cite{errington_2001,shi_2018} While translational order may be outside the scope of $E$, extending it to account for bond length information may, nevertheless, lead to a richer characterization of structure.

Uniquely among local structural indicators, $E$ shows both the qualities of an order parameter and a classifier. Remarkably, it does not do so at the expense of conceptual complexity, being one of the simplest quantities suited to local structural indication. One important question that goes largely unaddressed by improvements in our knowledge of the local structure of physical systems is how (and to what extent) larger-scale structural features can be inferred from local ones. The statistical properties of quantities such as $E$ may offer a path in this direction. Finally, while this work considers $E$ in the analysis of physical systems, such as crystals\cite{stukowski_2012} and liquids and glasses,\cite{tanaka_2019} it may also be suitable for studying particle-based systems in other areas, such as biology.\cite{da-fontoura-costa_2006, gibson_2009, jiao_2014}

\begin{acknowledgments}
The authors would like to thank Al{\'a}n Aspuru-Guzik, Chandra Veer Singh, and Zhirui Wang for their insightful discussions and two anonymous reviewers for their valuable suggestions.
\end{acknowledgments}

\bibliography{main.bib}

\appendix

\section{The robustified Voronoi neighborhood}
\label{appx:formal_demarcation}
Given a system $S$ of particles $p$ in $D$-dimensional Euclidean space $\mathbb{R}^D$, let $\{\varepsilon_{i,p} : i \in \mathbb{Z}^+, p \in S \}$ be a family of independent and identically distributed random vectors with uncorrelated Gaussian-distributed components of mean $\mu = 0$ and scale ${\sigma \ll \langle r_p \rangle}$, the average nearest-neighbor distance over all $p \in S$. Call a map $f_i : S \to \mathbb{R}^D$ defined by
\begin{equation}
    f_{i}(p) = p + \varepsilon_{i, p}
\end{equation}
a \textit{perturbation}. Then, the \textit{robustified Voronoi neighborhood} $\Vcal^*(p)$ of a particle $p$ is the set of all particles $q$ in the naive neighborhood $\Ncal_{\tau}(p)$ whose image under the perturbation $f_i$ is a member of the Voronoi neighborhood $\Vcal\big[f_i(p)\big]$ with probability greater than $0.5$. More formally,
\begin{equation}
    \Vcal^*(p) := \big\{ q \in \Ncal_{\tau}(p) :
     \Pr\left\{f_i(q) \in \Vcal\big[f_i(p)\big] \right\} > 0.5  \big\}.
\end{equation}

This is equivalent to a majority voting scheme wherein, for each $i$, the membership of perturbed particle $f_i(q)$ to the conventional Voronoi neighborhood of perturbed particle $f_i(p)$ constitutes one vote toward the membership of unperturbed particle $q$ to the robustified Voronoi neighborhood of unperturbed particle $p$. We restrict $\Vcal^*(p)$ to be a subset of $\Ncal_{\tau}(p)$ in order to prevent the unnatural result in which $p$ is assigned neighbors that are much further away from it than its nearest neighbor.

\section{Membership probability}
\label{appx:membership_prob}
The definition of the robustified Voronoi neighborhood $\mathcal{V}^*$ refers to the  probability that a perturbed naive neighbor $f_i(q), q \in \Ncal(p)$ of a particle $p$ is a member of its post-perturbation Voronoi neighborhood $\Vcal[f_i(p)]$. By the law of large numbers, this probability can be written as follows:
\begin{equation}
    \Pr\big\{f_i(q) \in \Vcal\big[f_i(p)\big] \big\} =  \!\! \lim_{M \to \infty} \frac{1}{M}{\sum\limits_{i = 1}^{M} \openone_{\Vcal\left[f_i(p)\right]}\big[f_i(q)\big]} ,
    \label{eq:lln}
\end{equation}
where $\openone$ denotes the {indicator function}. We evaluate the right-hand side of this equation numerically through the following Monte Carlo method:
\begin{enumerate}
    \item For each particle $p$, draw three samples from a Gaussian distribution with mean $0$ and scale $\sigma$ and add these to its xyz coordinates. These displaced particles comprise the perturbed system.
    \item Compute the Voronoi diagram of the perturbed system and store the adjacency matrix $A^{(i)}$ that it implies.
    \item Repeat the above steps $M$ times and take the elementwise average of the resulting $M$ adjacency matrices, \[A_{qp} := \avg{ A^{(i)}_{qp} } .\]
    \item Round each element of the resulting matrix $A$. 
    For sufficiently large $M$, the rounded elements of this matrix correspond to membership probabilities,
    \[\Pr\big\{f_i(q) \in \Vcal\big[f_i(p)\big] \big\} = \avg{A_{qp}} .\]
\end{enumerate}

\section{Error with respect to a commonly encountered geometry}
\label{appx:neighborhood_error}

Let $\Theta$ denote the bond angles of the particle under study and $\Phi$ the bond angles of a given commonly encountered geometry (CEG). Define the \textit{root-mean-square error} $\rho$ of $\Theta$ with respect to $\Phi$ by
\begin{equation}
\rho \! = \!
    \begin{cases}
    \! \sqrt{{\left\langle\min\limits_{\varphi\in\Phi}(\theta-\varphi)^2\right\rangle}_{\!\! \theta}} & \text{if} \enspace \{\argmin\limits_{\varphi\in\Phi} (\theta-\varphi)^2 :  \theta\! \in\! \Theta\} \! = \! \Phi, \\
    \infty &\text{otherwise}.
    \end{cases}
    \label{eq:error_model}
\end{equation}
The second case in this definition ensures that the error is finite only if every angle in $\Phi$ is the closest to at least one angle in $\Theta$.

An issue with raw $\rho$ is that it is biased toward CEGs with more angles (it reports lower errors for such geometries). To correct for this bias, we adjust $\rho_i$ with respect to each CEG $i$ by a correction factor $c_i>0$ as follows:
\begin{equation}
    \rho^*_i = c_i \rho_i.
\end{equation}
Correction factors $c_i$ are chosen so that the expected error of a random geometry with respect to any CEG is the same. Since the cutoff for $\rho$  discussed in Sec. IV. B is chosen with respect to the fraction of FCC particles, we pick FCC as our reference geometry in choosing $c_i$. In other words,
\begin{equation}
    c_i := \left\langle\frac{ \rho_\text{FCC} }{\rho_{i}}\right\rangle,
\end{equation}
where $\rho_i$ are computed from the first case of Eq. (\ref{eq:error_model}).

Random neighborhoods are generated by performing k-means clustering on a set of sample points uniformly distributed over a spherical shell. The outer to inner radius ratio of the shell is set to $1.3$, approximately corresponding to the width of the first peak of the radial distribution function for RCP systems. The number of clusters is set to $14$, which is the closest integer to the average number of Voronoi cell facets in RCP systems.\cite{finney_2013}

\section{Estimator of bond angle count}
\label{appx:bond_angle_count_estimate}

To devise an estimator of bond angle count, we begin by considering the properties that such an estimator must satisfy. Visibly, the true number of different bond angles $m$ in a geometry is a nonnegative integer bounded from above by the number of bonds pairs. For the estimator $\hat{m}$, we relax the integer requirement to alleviate the chance of false positives on a CEG. This leaves us with the following two properties as a starting point:
\begin{align*}
\text{I.} \quad  \null &\hat{m}(k)\geq 0 \enspace \text{for} \enspace k>1          &\null \text{(nonnegativity),} \\
\text{II.} \quad \null &\hat{m}(k)\leq (k^2-k)/2   &\null \text{(upper bound).} \\
\end{align*}

CEGs exhibit an unusually large number of symmetries compared to arbitrary geometries of the same $k$. On the premise that symmetries in the neighborhood of a particle reduce its number of bond angles, it is unlikely for an unrecognized geometry with a given $k$ to have less angles than a CEG of the same $k$. We therefore also stipulate the following:
\begin{align*}
\text{III.} \quad  \null &\hat{m}(k) \geq \max\limits_{i \in G(k)} m_i  &\null \text{(lower bound)},
\end{align*}
where $i$ denotes a CEG, $G(k)$ denotes the set of all $k$-coordinate CEGs, and $m_i$ denotes the true number of different bond angles for CEG $i$. As a final consideration, it is clear that the the the number of different angles possible increases with the number of bonds. We hence require the following:
\begin{align*}
\text{IV.} \quad  \null &\hat{m}(k+1) \geq \hat{m}(k)           &\null \text{(increasing monotonicity)}.
\end{align*}

Having established a set of properties that we desire from the estimator, we now choose a simple functional form that is able to satisfy them,
\begin{equation}
    \label{eq:functional_form}
    \hat{m}(k) = b - a \exp( -k / \gamma),
\end{equation}
where $a, b, \gamma > 0$. Functions of this form are monotonically increasing in $k$ for all positive $a$, $b$, and $\gamma$. Property IV is thereby automatically satisfied. Moreover, it is visible that for any choice of $\gamma$, one can select $a$ and $b$ to fit
\begin{equation}
    \label{eq:points}
    \begin{aligned}
        \hat{m}(2)& = 1, \\
        \hat{m}(12)& = 7;
    \end{aligned}
\end{equation}
these being the points of smallest and largest known $m$, respectively. Thus, Property I is also automatically satisfied and it remains only to choose $\gamma$ to satisfy Property II and III. 

We find that Property II is violated for $\gamma \lessapprox 2.4$ and that Property III is violated for $\gamma \gtrapprox 4.3$. We take the average of these values, $\gamma = 3.4$, for which both properties are satisfied. The values of $a$ and $b$ that fit Eq. (\ref{eq:points}) for $\gamma = 3.4$ are as follows:
\begin{equation}
    \begin{aligned}
        a &= 11.407186472007339,& \\
        b &= \phantom{1}7.334483343628186.&
    \end{aligned}
\end{equation}


\end{document}